\documentclass[twocolumn,showpacs,amsmath,amssymb]{revtex4}

\usepackage{graphicx}% Include figure files
\usepackage{dcolumn}% Align table columns on decimal point
\usepackage{bm}% bold math
\usepackage{psfig}  
\usepackage{epsfig}

\begin{document}

\title{Jet quenching at intermediate RHIC energies}

\author{Ivan Vitev}
\email{ivitev@iastate.edu}

\affiliation{Department of Physics and Astronomy, 
Iowa State University, Ames, IA 50011, USA }

\begin{abstract}
The final state energy loss of fast partons penetrating a 
longitudinally expanding quark-gluon plasma of effective gluon 
rapidity density $dN^{g}/dy=650-800$ is evaluated and incorporated  
together with the multiple initial state Cronin scattering in 
the lowest order perturbative QCD hadron production formalism. 
Predictions for the neutral pion attenuation in central $Au+Au$ 
collisions at the intermediate RHIC energy of $\sqrt{s_{NN}}=62$~GeV 
relative to the binary collision scaled $p+p$ result are given.
The quenching is found to be a factor of $2-3$ with a moderate
transverse momentum dependence and the attenuation of the  
away side di-hadron correlation function is estimated to be $3-5$ fold. 
\end{abstract}
                                               

\maketitle

%%%%%%%%%%%%%%%%%%%%%%%%%%%%%%%%%%%%%%%%%%%%%%%%%%%%%%%%%%%%%%%%%%%

\section{Introduction}

Recent combined experimental measurements of the nuclear 
modification to the moderate and large transverse momentum 
hadron production in $Au+Au$~\cite{Adler:2003qi,Adler:2003au} 
and $d+Au$~\cite{Arsene:2003yk} reactions have provided 
strong evidence in support of the dominance of 
multiple final state interactions~\cite{Gyulassy:2003mc}  
over the initial state Cronin scattering~\cite{Zhang:2001ce} 
and possible nuclear wavefunction effects~\cite{Kharzeev:2002pc} 
in relativistic heavy ion collisions.  
These findings pave the way for detailed studies of derivative
jet quenching observables~\cite{Gyulassy:2000gk,Vitev:2004bh} such 
as the high-$p_T$ azimuthal anisotropy~\cite{Adler:2002ct},
the broadening and disappearance of di-jet 
correlations~\cite{Adler:2002tq}, extensions of the 
correlation analysis with respect to the reaction 
plane~\cite{Bielcikova:2003ku}  and  possibilities  for full jet 
and lost energy reconstruction~\cite{Wang:2004kf}.
The theory and phenomenology of multiparton dynamics in 
ultra-dense nuclear matter will, however, remain incomplete without 
a thorough investigation of the center of mass energy and system 
size dependence~\cite{Vitev:2002pf,Drees:2003zh} 
of medium induced non-Abelian gluon bremsstrahlung
and the corresponding hadron attenuation.
Possible onset of pion suppression at the 
SPS energy of $\sqrt{s_{NN}}=17$~GeV has been 
recently discussed~\cite{d'Enterria:2004ig}. 
The intermediate $\sqrt{s_{NN}}=62$~GeV RHIC run is the next 
critical step in mapping out the elastic and inelastic 
scattering properties of a quark-gluon plasma 
via jet tomography~\cite{Vitev:2004bh,Gyulassy:2004vg}.

The magnitude of the energy loss driven nuclear quenching of 
moderate and high $p_T$ pions is controlled by  the soft 
parton rapidity density~\cite{Vitev:2004bh,Gyulassy:2004vg}. To 
relate the experimentally measured 
$dN^{ch}/d\eta$~\cite{Back:2001ae} to the effective 
$dN^g/dy$ we use  $|d \eta /dy| \approx 1.2 $ at $\eta=y=0$, 
ignoring the $\sqrt{s_{NN}}$ dependent  changes in  particle 
composition. The estimated effective 
\begin{equation}
\frac{dN^g}{dy} \approx \frac{3}{2}\,
\left| \frac{d\eta}{d y} \right| \, \frac{dN^{ch}}{d \eta}
\label{rap-dens}
\end{equation}
follows from the isospin symmetry of strong interactions 
and the approximate parton-hadron duality~\cite{Gell-Mann:nj}.
Straightforward  application of Eq.~(\ref{rap-dens}) for 
central, $N_{part} = 340$, $Au+Au$ collisions and $dN^{ch}/d\eta$
constrained from the data~\cite{Back:2001ae} 
yields  $dN^g/{dy} \approx 550, \, 850, \, 1150$ at
SPS, the intermediate and maximum RHIC energies, 
respectively. Such rapidity densities will likely be compatible
with a soft participant scaling phenomenology~\cite{Kharzeev:2000ph}. 
The predicted $\sqrt{s_{NN}}=200$~GeV pion quenching~\cite{Vitev:2002pf}  
was found to be in good  agreement with the experimental 
measurements~\cite{Adler:2003qi} but the original
SPS $\pi^0$ data~\cite{Aggarwal:2001gn} would tend to disfavor 
parton energy loss in variance with the theoretical 
expectations~\cite{Gyulassy:2003mc}. A more recent 
analysis of the low energy $p+p$ baseline cross 
section~\cite{d'Enterria:2004ig} shows that the 
WA98~\cite{Aggarwal:2001gn} and CERES~\cite{Agakishiev:2000bt} 
data are not inconsistent with jet  quenching calculations 
with $dN^g/dy=400$ when the
Cronin effect is taken into account~\cite{Vitev:2002pf}.    
The extracted effective gluon rapidity density, however, still 
falls short of the expectation from the measured 
hadron multiplicities~\cite{Back:2001ae}.  
This deviation can be related to the uncertainties in the 
perturbative calculation in a theory with strong coupling  
and a non-negligible quark contribution~\cite{quarks} to the bulk 
soft partons at $\sqrt{s_{NN}}=17$~GeV at midrapidity.
Therefore, the importance of studying the sensitivity of the 
observable spectral modification to a range of parton densities 
at a fixed center of mass energy should not be underestimated 
in theoretical calculations.

The purpose of this letter is to investigate the interplay 
of the initial state multiple Cronin scattering and the 
final state energy loss in a thermalized QCD media 
with effective $dN^g/dy = 650-800$  in moderate and large
transverse momentum hadron production.  
Section~II outlines the method for evaluating the 
medium induced gluon bremsstrahlung  off fast partons in dense, 
dynamically expanding and finite nuclear matter. 
Section~III presents the calculated 
depletion of pion multiplicities in central $Au+Au$ 
reactions at $\sqrt{s_{NN}}=62$~GeV relative to the binary 
collision scaled $p+p$ baseline. Summary and discussion
is given in Section~IV.

\section{Medium induced gluon bremsstrahlung in finite 
dynamical plasmas}

\begin{figure*}[t!]
\begin{center} 
\psfig{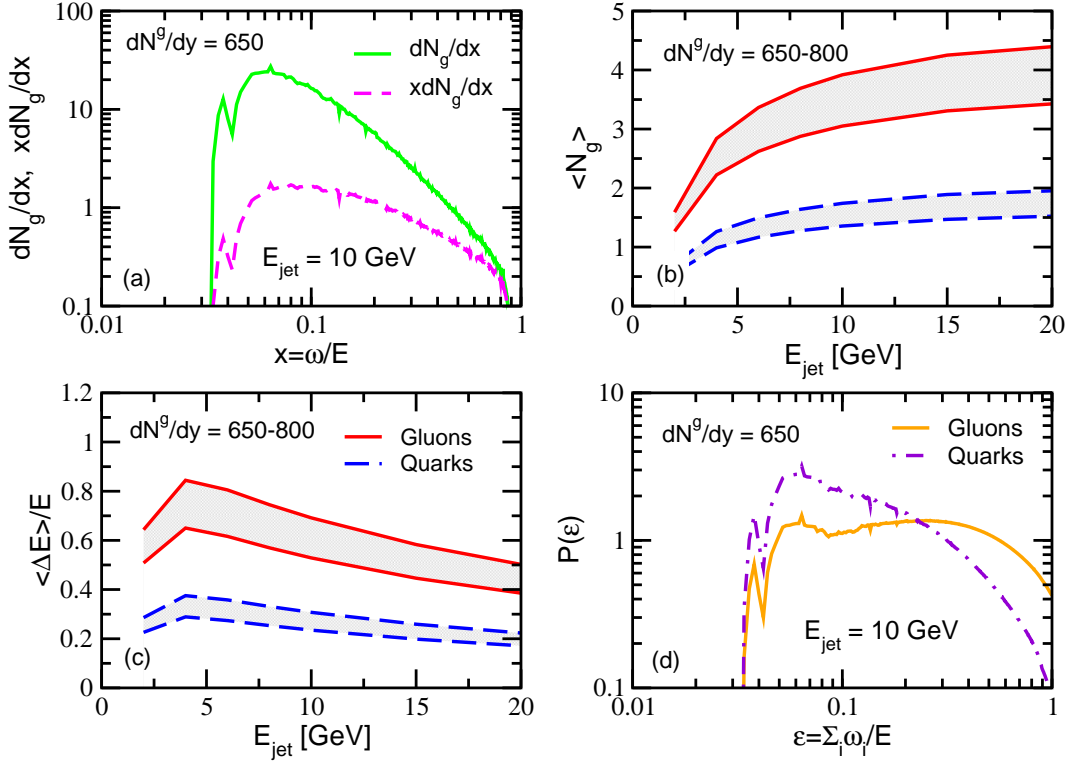}
\vspace*{0.in}
\caption{(a) Medium induced single inclusive 
gluon spectrum $dN_g/dx$ and  fractional intensity 
$xdN_g/dx$ versus  $x \approx \omega/E$ for a $E=10$~GeV gluon jet. 
(b) Mean medium-induced gluon number $\langle N_g\rangle$ per jet 
for initial effective soft parton rapidity density 
$dN^g/dy = 650-800$.
(c) Corresponding mean fractional energy loss 
$\langle \Delta E  \rangle / E$. 
(d) Probability distribution $P(\epsilon)$ of the fractional 
energy loss  of $E=10$~GeV quarks and gluons via multiple 
independent  gluon bremsstrahlung, 
$\epsilon = \sum_i \omega_i / E$. }
\label{fig1:F-rule}
\end{center} 
\end{figure*}

The full solution for the medium induced gluon radiation
off jets produced in a hard collisions at early 
times $\tau_{jet} \simeq 1/E$ inside a  nuclear  
medium  of length $L$  can be obtained to all orders  
in the correlations  between the  multiple  scattering  centers  
via the reaction operator approach~\cite{Gyulassy:2000er}.
Other existing techniques have been reviewed in~\cite{Gyulassy:2003mc}.
The double differential bremsstrahlung  intensity 
for gluons with momentum $k=[xp^+, {\bf k}^2 / xp^+,{\bf k}]$  
resulting from the sequential interactions 
of a fast parton with momentum $p=[p^+, 0,{\bf 0}]$ 
can be written as 
\begin{widetext}
\begin{eqnarray}
 x\frac{dN_g}{dx\, d^2 {\bf k}}  &=&
\sum\limits_{n=1}^\infty  x\frac{dN_g^{(n)}}{dx\, d^2 {\bf k}}   
 = \sum\limits_{n=1}^{\infty}  \frac{C_R \alpha_s}{\pi^2} 
 \; \prod_{i=1}^n \;\int_0^{L-\sum_{a=1}^{i-1} \Delta z_a } 
 \frac{d \Delta z_i }{\lambda_g(i)} 
 \,  \int   d^2{\bf q}_{i} \, 
\left[  \sigma_{el}^{-1}(i)\frac{d \sigma_{el}(i)}{d^2 {\bf q}_i}
  - \delta^2({\bf q}_{i}) \right]  \,  \nonumber \\[1.ex] 
&\;& \times 
\left( -2\,{\bf C}_{(1, \cdots ,n)} \cdot 
\sum_{m=1}^n {\bf B}_{(m+1, \cdots ,n)(m, \cdots, n)} 
\left[ \cos \left (
\, \sum_{k=2}^m \omega_{(k,\cdots,n)} \Delta z_k \right)
-   \cos \left (\, \sum_{k=1}^m \omega_{(k,\cdots,n)} \Delta z_k \right)
\right]\; \right) \;, \quad \qquad  
\label{difdistro} 
\end{eqnarray}
\end{widetext}
where $\sum_2^1 \equiv 0$ is understood. In the small angle 
eikonal limit 
$x=k^+/p^+ \approx \omega/E$. In Eq.~(\ref{difdistro})  
the color current propagators are denoted by
\begin{eqnarray}
{\bf C}_{(m, \cdots ,n)} &=&  \frac{1}{2} \nabla_{{\bf k}} 
\ln \, ({\bf k} - {\bf q}_m - \cdots  - {\bf q}_n )^2 \nonumber \\  
{\bf B}_{(m+1, \cdots ,n)(m, \cdots, n)} &=&  
{\bf C}_{(m+1, \cdots ,n)} - {\bf C}_{(m, \cdots ,n)} \;\;. 
\label{props}
\end{eqnarray}
The momentum transfers ${\bf q}_i$ are distributed according to 
a normalized elastic differential cross section,  
\begin{equation}
\sigma_{el}(i)^{-1}\frac{d \sigma_{el}(i)}{d^2 {\bf q}_i}  
= \frac{\mu^2(i)}{\pi({\bf q}_i^2+\mu^2(i))^2} \; ,
\label{GWmodel}
\end{equation}
which models scattering  by  soft partons with a thermally 
generated Debye screening mass $\mu(i)$. 
It has been shown~\cite{Gyulassy:2000er,Wiedemann:2000za} via the 
cancellation of direct and virtual diagrams that in the 
eikonal limit only the gluon mean free path $\lambda_g(i)$ 
enters the medium-induced  bremsstrahlung spectrum in  
Eq.~(\ref{difdistro}). For gluon dominated bulk soft matter 
$\sigma_{el}(i)  \approx  \frac{9}{2} \pi \alpha_s^2/\mu^2(i)$ 
and  $\lambda_g(i)=1/\sigma_{el}(i)\rho(i)$.  
The characteristic path length dependence of the non-Abelian energy 
loss in Eq.~(\ref{difdistro}) comes from the
interference phases and is differentially controlled
by the inverse formation times,    
\begin{equation}
\omega_{(m,\cdots,n)}  = 
\frac{({\bf k} - {\bf q}_m - \cdots  - {\bf q}_n )^2}{2 x E}  \;, 
\label{ftimes}
\end{equation}
and the separations of the subsequent scattering centers  
$ \Delta z_k = z_k - z_{k-1} $. It is the non-Abelian 
analogue of the Landau-Pomeranchuk-Migdal destructive
interference effect in QED~\cite{Landau:um}.

For the case of local thermal equilibrium we relate all 
dimensional scales in the problem to the temperature of the 
medium $T(i)$, for example, $\mu^2(i) = 4 \pi \alpha_s T(i)^2$ 
and the elastic scattering cross section indicated above.  
The gluon radiative spectrum is evaluated  numerically 
in an ideal  1+1D Bjorken expanding plasma~\cite{Bjorken:1982qr},  
$\rho(i) = \rho_0 (\tau_0 / \tau_i)^\alpha$, $\epsilon_i = 
\epsilon_0 (\tau_0 / \tau_i)^{\alpha +v_s^2}$ with $ \alpha = 1$
and $v_s$ being the speed of sound. 
The scaling with proper time of the temperature, 
the Debye screening mass $\mu(i)$ and the gluon mean free 
path $\lambda_g(i)$ naturally follow. To leading power, the initial 
equilibration  time $\tau_0$  cancels in the evaluation of 
the bremsstrahlung  integrals~\cite{Gyulassy:2003mc} since 
\begin{equation} 
\int^L_0 d\tau_i \; \rho(i) \tau_i = 
\frac{1}{A_\perp} \frac{dN^g}{dy} \; L  \;\;.
\label{BJ}
\end{equation} 
In Eq.~(\ref{BJ}) $A_\perp$ is the transverse size of the 
medium and $dN^g/dy$ is the relevant physical quantity 
that controls the attenuation of the final state partonic flux.

Transverse, $\beta_T \neq 0$, expansion was not explicitly included 
in the energy loss calculation. Hence, we first clarify its impact
on the medium induced non-Abelian bremsstrahlung. Numerically, 
the contributions to  the radiative spectrum in  Eq.~(\ref{difdistro})  
have to be evaluated in the background of the dynamically 
evolving soft parton multiplicity. In practice, for
realistic 3+1D hydrodynamic simulation only the dominant 
$n=1$ term~\cite{Gyulassy:2000er} has been considered in 
the mean energy loss approximation~\cite{Hirano:2003hq}.
Analytic treatments of transverse expansion must therefore 
provide important guidance to its effect on the radiative 
spectra and $\langle \Delta E \rangle$.

The simplest approach to $\beta_T \neq 0$ would be to modify 
the power  $\alpha$ of the 1+1D Bjorken case to emulate 3+1D 
dynamics.  Indeed, beyond leading power, Eq.~(\ref{BJ}), 
$\alpha_s^2(\mu)$ effectively {\em lowers} the value of Bjorken 
$\alpha$ relative to the naive fixed coupling result. It has also 
been argued~\cite{Bjorken:1982qr} that if the transport properties 
of the medium are related to the energy density, deviations from 
the ideal plasma limit would lead to $\alpha < 1$
via $v_s^2 = (3+\Delta)^{-1}$ with $\Delta = \frac{165}{8} 
\left( \frac{\alpha_s}{\pi} \right)^2 + \cdots $ .
Technically, if $\alpha > 1$ the density integrals that 
control the energy loss can still be defined~\cite{Salgado:2003gb}.  
However, for a medium of fixed size $L$ where simple analytic 
results can be obtained, Eq.~(\ref{BJ}), or even for a 
spatially varying density profile $\rho_0 = \rho_0({\bf x}_T)$, 
such  physical picture  corresponds to a superluminal 
Bjorken expansion. In this scenario, the dilution of the quark-gluon
plasma  from the transverse flow of the soft partons 
is modeled by streaming along the collision axis. 
As a result, the partons are lost as potential 
scatterers for the hard jets that escape the plasma at $y=0$ 
and the energy loss is reduced. Another consequence 
of choosing  $\alpha > 1$  is the {\em decrease} of the  
accumulated  transverse momentum  $\langle k_T^2 \rangle \propto 
\int_{\tau_0}^L \rho_0 (\tau_0 / \tau)^\alpha d \tau $ with the size 
of the medium $L$ or, in realistic $A+A$ collisions, significantly 
lowering the dependence of $\langle k_T^2 \rangle$ on centrality. 
This contradicts the measured steady growth of the di-jet 
acoplanarity  versus $N_{part}$~\cite{Adler:2002tq} which even 
exceeds the theoretical expectations for the Bjorken $\alpha = 1$ 
case~\cite{Vitev:2004bh}.

Varying $\alpha$ and attributing the  deviation from unity, 
$\alpha - 1$, to a model of spatially non-uniform  
medium leads to power law behavior $\rho ( {\bf x}_T ) 
=  \rho_0 |{\bf x}_{T } - {\bf x}_{T 0} |^{\alpha - 1}$  
rather than the smooth Woods-Saxon dependence of the 
nuclear matter density. Additionally, there is a large ambiguity 
in the choice of $\alpha$ since as a function of the azimuthal
angle $\phi$ of the jet propagation relative to the reaction plane 
the spatial density can either increase or decrease.                
We thus conclude that $ \alpha > 1$ is not a good emulation
of the realistic $\beta_T \neq 0$  expansion  and  that the
geometry profile and the soft parton dynamics should 
not be substituted for each other.

An analytically tractable approximation that illustrates 
the effects of transverse expansion has been developed 
in~\cite{Gyulassy:2001kr}. In this scenario, for a 
medium of mean density $\rho_0$ and mean radius $R = L$ 
\begin{equation}          
\rho (\tau) =  \frac{1}{\pi}  \frac{dN^g}{dy } \frac{1}{\tau}  
\frac{1}{(L + \beta_T \tau)^2}  \;.
\label{3d}
\end{equation}          
The additional dilution relative to the 1+1D Bjorken case  
now generated by the transverse motion of the soft partons
and the increasing transverse size $A_\perp$ of the system. 
For a discussion of non-central 
collisions, $R_x \neq R_y$ and  $\beta_{T\,x} \neq \beta_{T\,y}$, 
see~\cite{Gyulassy:2001kr}. It follows from Eq.~(\ref{3d}) that 
the propagating jets interact with the bulk matter at 
midrapidity over an 
increased time period $\Delta \tau \simeq L/(1-\beta_T)  > L$ 
that largely compensates for the rarefaction of the medium.
It has been demonstrated~\cite{Gyulassy:2001kr} that for 
small and moderate expansion velocities $\beta_T$ the 1+3D 
angular ($\phi$) averaged energy loss approximates 
well the 1+1D result. Logarithmic corrections arise only in 
the $\beta_T \rightarrow 1$ limit. It should  now  be 
physically intuitive why the Bjorken expansion is a good 
approximation in the calculation of quenching in the single 
and double inclusive hadron spectra.

An important aspect of the application of the theory
of medium induced non-Abelian bremsstrahlung is the inclusion 
of finite kinematic bounds. These are particularly 
relevant at RHIC and SPS energies over the full 
accessible $p_T$ range. Additional details on the 
evaluation of the radiative energy loss are given 
in~\cite{Gyulassy:2000er}.

Figure~1(a) shows the differential gluon spectrum 
$dN_g/dx$ and the fractional radiation intensity 
$xdN_g/dx$ computed to 3-rd order in opacity ($n=1,2,3$) 
from Eq.~(\ref{difdistro}) for a 10~GeV gluon jet. 
The fluctuations in the calculation reflect the 
numerical accuracy of the cancellation between the 
propagator poles, Eq.~(\ref{props}), and the interference 
phases, Eq.~(\ref{ftimes}), and come predominantly from 
higher orders in opacity. These, however, do not affect 
the evaluation of the nuclear modification factor. 
The jet quenching strength is largely 
set by the $n=1$ term~\cite{Gyulassy:2000er}.

In a thermalized medium the plasmon frequency 
$\omega_{pl}(i)\sim \mu(i)$  regulates the infrared 
modes and the medium induced  bremsstrahlung  results in  
a few semihard gluons. Numerical results for quark jets are 
not given since they differ by a 
simple $C_F/C_A = 4/9$ color factor. 
Figures~1(b) and 1(c) show the mean induced gluon   
number $\langle N_g \rangle$ and the 
mean fractional energy  loss $\langle \Delta E \rangle / E$ 
calculated directly from Eq.~(\ref{difdistro}) 
for a range of soft parton  rapidity densities $dN^g/dy = 650 - 800$. 
Comparing the shape of $\langle \Delta E \rangle / E$
in Figure~1(c) to the naive analytic expectation in the  infinite 
kinematic limit  
\begin{equation}
\frac{\langle \Delta E \rangle}{E}
\approx \frac{ 9 C_R \pi \alpha_s^3  }{4}
 \frac{1}{A_\perp} \frac{dN^{g}}{dy} \, L 
 \; \frac{1}{E} \, \ln   \frac{2 E}{\mu^2 L}  +  \cdots 
\label{analyt-de}
\end{equation}
and observing the deviation from the hyperbolic $1/E$ dependence 
it is easy to recognize the need for a careful treatment 
of phase space. It is  also instructive  to note that 
the average energy  per gluon  $\langle \Delta E \rangle/ 
\langle N_g \rangle  \simeq  0.8-1.9\;{\rm GeV}$, which 
implies that the radiative quanta might be experimentally observable with
finite $p_T$ cuts~\cite{Vitev:2004bh,Wang:2004kf}.

Applications that extend beyond the mean energy loss
and invoke a probabilistic treatment with multiple gluon 
fluctuations require additional assumptions~\cite{Baier:2001yt}. 
So far even the case of two gluon emission with scattering has 
not been calculated. The Poisson approximation assumes 
independent radiation 
and equivalence of the single  inclusive and  the single 
exclusive gluon spectrum.  Finite kinematics alone is 
guaranteed to violate this ansatz and so is angular ordering. 
Nevertheless, the usefulness of such probabilistic treatment 
is in allowing the system to maximize the observable 
cross section and correspondingly minimize, to the extent 
to which this is possible, the effect of the non-Abelian 
medium-induced bremsstrahlung.

The probability $P(\epsilon)$ for fractional energy loss  
$\epsilon = \sum_i \omega_i / E$ due to multiple gluon emission 
is given in Figure~1(d). The $\epsilon = 0$ bin contribution 
$\delta(\epsilon)e^{-{\langle N_g \rangle}}$ is not shown. Details 
on the calculation of $P(\epsilon)$ are given in~\cite{Baier:2001yt}, 
where a fast iterative procedure for its evaluation form the gluon 
radiative spectrum  $dN_g/dx$ was developed. If a non-zero 
fraction of the probability $P(\epsilon)$ falls in the
$\epsilon > 1$ region, it is uniformly redistributed in the
physical $\epsilon \in [0,1]$ interval.   
This approach ensures that $\langle \Delta E \rangle / E \leq 1$ 
in the large energy loss limit and provides corrections relative 
to its direct evaluation from Eq.~(\ref{difdistro}) that 
are  different for quarks and gluons.   
The properties of $P(\epsilon)$ can be summarized by the 
normalization of the first two $\epsilon^k$, $k=0,1$ moments:
\begin{equation}
\int_0^1 P(\epsilon) \; d\epsilon = 1, \quad 
\int_0^1 \epsilon \, P(\epsilon) \; d \epsilon =
 \frac{\langle \Delta E \rangle}{E}  \;\;.
\end{equation} 
We note the distinct difference in the manifestation of the large 
energy loss relative to the single inclusive spectra. In the former 
case the overall scale  changes and in the probabilistic 
interpretation  the distribution is shifted toward large 
$\epsilon$ values, see Figure~1(d).

\section{Phenomenological application at intermediate 
energies}

Dynamical  nuclear  effects  in $p+A$ and $A+A$ reactions     
are most readily detectable through  the nuclear modification  ratio 
\begin{equation}         
R^{h_1}_{AA}({\bf p}_{T1})  = 
\frac{d N^{AA}({\bf b})}{dy_1d^2{\bf p}_{T1}} \big/ 
\frac{ T_{AA}({\bf b})\; d \sigma^{pp}}{dy_1d^2{\bf p}_{T1}}
   \;\;   \quad   {\rm in} \;  \; A+A  \;  ,  
\label{geomfact} 
\end{equation}          
where $T_{AA}({\bf b}) = \int  d^2{\bf r} \, 
T_A({\bf r})T_B({\bf r}-{\bf b})$ is calculable  in terms of the 
nuclear thickness functions $T_A({\bf r})=\int dz \,\rho_A({\bf r},z)$.
In $R^{h_1}_{AA}({\bf p}_{T1})$ the uncertainty  associated with the
 next-to leading order factor,  $K_{NLO}$, drops out.

\begin{figure*}[t!]
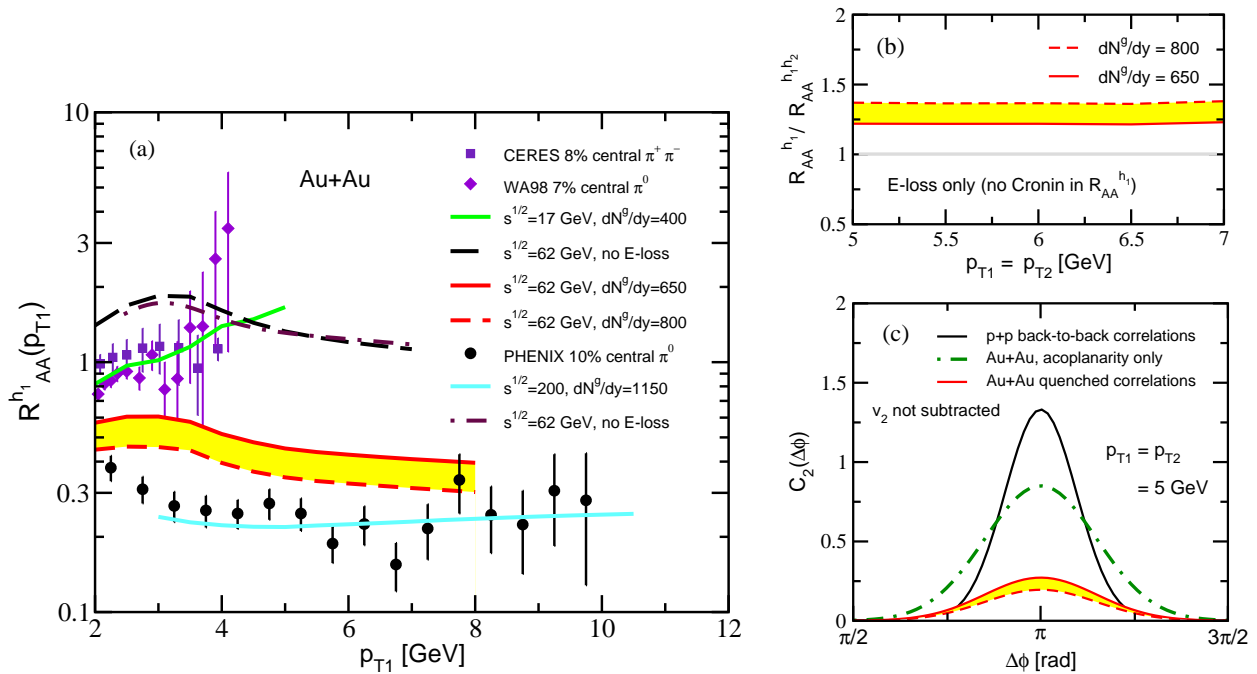

\begin{center} 
\psfig{file=Fig2.NEWa.eps,height=3.in,angle=0}
\hspace{.1in}
\psfig{file=Fig2.NEWb.eps,height=3.5in,angle=0}
\vspace*{0.in}
\caption{(a) Calculated nuclear modification factor  
$R_{AA}({\bf p}_{T1})$ versus the  center of mass energy 
$\sqrt{s_{NN}}=17 , \, 62, \, 200$~GeV for central $Au+Au$ collisions.
The yellow band represents the perturbative QCD expectation for the 
depletion of the neutral pion multiplicity
at $\sqrt{s_{NN}}=62$~GeV. Enhancement, arising from 
transverse momentum diffusion in cold nuclear matter 
without final state energy loss
is given for comparison. PHENIX~\cite{Adler:2003au}, 
WA98~\cite{Aggarwal:2001gn} and CERES~\cite{Agakishiev:2000bt}  
data as presented in~\cite{d'Enterria:2004ig} are shown.
(b) The ratio of the attenuation of the single and 
double inclusive $\pi^0$ cross sections 
$ R^{h_1}_{AA}(p_{T1}) / R^{h_1h_2}_{AA}(p_{T_1}= p_{T_2})$ 
in central Au+Au collisions at $\sqrt{s_{NN}}=62$~GeV
and  (c) the  manifestation of the double inclusive hadron suppression 
in the quenching of the away-side di-hadron correlation function 
$C_2(\Delta \phi)$~\cite{Qiu:2004da}. }
\label{fig2}
\end{center} 
\end{figure*}

The standard lowest order perturbative expression for the 
hadron multiplicity  in  $A+A$ reactions including the effects 
of multiple initial state 
scattering and the final state energy loss 
reads~\cite{Brock:1993sz,Vitev:2002pf}: 
\begin{widetext}
\begin{eqnarray}
\frac{1}{T_{AA}({\bf b})}
\frac{d N^\pi_{AA}({\bf b})}{dy_1d^2{\bf p}_{T1}} &=& K_{NLO}   
\sum_{abcd} \int_{x_{a\, \min}}^1 \!\!\! dx_a  \int_{x_{b\, \min}}^1
\!\!\! dx_b  \int  d^2 {{\bf k}}_a  \int d^2{{\bf k}}_b \; \;  
g({{\bf k}}_a) g({{\bf k}}_b) \;
 f_{a/A}(x_a,Q^2) f_{b/A}(x_b,Q^2) \nonumber \\ 
&&\,  \times  \; S_{a/A}(x_a,Q^2)  S_{b/A}(x_b,Q^2)  
\; \frac{ \alpha_s^2}{\hat{s}^2} 
\, \big|\bar{M}_{ab\rightarrow cd}(\hat{s},\hat{t},\hat{u})\big|^2
\int_0^1 d \epsilon \, P(\epsilon) 
 \frac{z^*_c}{ z_c} \frac{D_{\pi/c}(z^*_c,{Q}_c^2)}{ z_c} \; . 
\label{pqcdLO} 
\end{eqnarray}
\end{widetext}
In Eq.~(\ref{pqcdLO}) $f_{a/A}(x,Q^2)$ are the isospin corrected 
($A=Z+N$) lowest order parton distribution 
functions~\cite{Gluck:1998xa},  $S_{a/A}(x,Q^2)$ 
is the leading twist shadowing parameterization~\cite{Eskola:1998df}, 
and $D(z)_{\pi/c}(z,Q^2)$ is the fragmentation function into 
pions~\cite{Binnewies:1994ju}.  
Vacuum and  medium-induced initial state parton broadening,  
$\langle {\bf k}_T^2 \rangle = \langle {\bf k}_T^2 \rangle_{pp} +  
\langle {\bf k}_T^2 \rangle_{nucl.} $, 
is incorporated  via a normalized Gaussian ${\bf k}_T$ smearing 
function~\cite{Brock:1993sz}.     
If the final state parton looses a fraction $\epsilon$  of its energy the 
correspondingly rescaled fragmentation momentum fraction reads
$z^* = z/( 1 -\epsilon)$.  For consistency the calculation is performed 
in the same way as in~\cite{Vitev:2002pf} where additional 
details can be found.

In the $p_T \leq 5$~GeV range, experimentally accessible at 
$\sqrt{s_{NN}}=62$~GeV, sizable 
non-perturbative effects in baryon production, manifest in
enhanced $p/\pi$ and $\Lambda/K$ ratios,  will likely 
be observed. Discussion  of the moderate $p_T$ baryon phenomenology 
is beyond the scope of this letter and details are given  
in~\cite{Vitev:2001zn}. In the limit of vanishing baryon masses, 
$m_B \rightarrow 0$, perturbative calculations of 
the nuclear modification factor $R^{h_1}_{AA}({\bf p}_{T1})$ yield results 
comparable to the one for neutral and charged pions. 
We finally note that the dominant contribution to the nuclear 
shadowing may come from enhanced dynamical power 
corrections~\cite{Qiu:xy,Qiu:2003vd}, resulting from the multiple 
initial and final state interactions of the partons on a nucleus. 
However, in the calculated $p_T$ range ($x \geq 0.8$, $Q^2 \geq 8$~GeV$^2$) 
their effect was found to be small.

Results form the perturbative calculation of the nuclear 
modification to the neutral and charged pion production in central 
$Au+Au$ collisions are given in the left hand side of 
Figure~2(a). At all energies there is a strong cancellation between 
the Cronin enhancement, which arises from the transverse momentum 
diffusion of fast partons in cold nuclear 
matter~\cite{Gyulassy:2000gk,Gyulassy:2000er}, 
and the subsequent inelastic jet attenuation in the  final state. 
This interplay is most pronounced at the low SPS $\sqrt{s_{NN}}=17$~GeV  
where, in the absence of quenching, the corresponding enhancement 
could reach a factor of 3-4~\cite{Vitev:2002pf}. With final state 
energy loss taken into account, $R^{h_1}_{AA}({\bf p}_{T1})$ is 
shown versus the reanalyzed~\cite{d'Enterria:2004ig}
WA98~\cite{Aggarwal:2001gn} and the CERES~\cite{Agakishiev:2000bt} 
$Pb+Pb$ and $Pb+Au$ data. We note that even the previously 
extracted nuclear modification at the SPS allows for inelastic final 
state interactions in a medium of $dN^g/dy=200$~\cite{Vitev:2002pf}.

At present, there are strong 
indications~\cite{Gyulassy:2004vg,Vitev:2004bh} 
that the highest nuclear matter density reached in the early stages  
($\tau_0=0.6$~fm)  of the $\sqrt{s_{NN}}=200$~GeV  
central $Au+Au$ collisions at RHIC 
may be on the order of 100 times cold nuclear matter 
density, $\epsilon_{cold}=0.14$~GeV/fm$^3$. 
Perturbative calculations with a corresponding 
$dN^g/dy=1150$ are compatible 
with the measured $\pi^0$ attenuation~\cite{Adler:2003qi},  
as illustrated in Figure~2(a). Quenching ratios of similar magnitude,
$R^{h_1}_{AA}({\bf p}_{T1}),\, R^{h_1}_{CP}({\bf p}_{T1}) \sim 0.2-0.25$,  
have also been predicted and measured  for inclusive charged hadrons 
at $p_T \geq 5$~GeV~\cite{Adler:2003qi}.  It is interesting to note 
that in spite of the very large initial energy density the 
nuclear attenuation is ``only'' a factor of $4-5$. The 
reason for this somewhat unintuitive result is 
the strong longitudinal expansion in the absence of which 
energetic jets would have been completely absorbed.

The yellow band represents a calculation of  the 
$\pi^0$ attenuation at $\sqrt{s_{NN}}=62$~GeV from  
Eqs.~(\ref{geomfact}), (\ref{pqcdLO}). The Cronin enhancement alone 
in central $Au+Au$ reactions was found to be still sizable with 
$R^{h_1}_{AA}({\bf p}_{T1})_{\max} = 1.8 - 2$ at $p_T = 3 - 3.5$~GeV and 
a subsequent decrease at higher transverse momenta. For comparison,
at $\sqrt{s_{NN}}=200$~GeV the Cronin maximum decreases with the 
center of mass energy and $R^{h_1}_{AA}({\bf p}_{T1})_{\max} = 1.7$. 
While such enhancement may naively appear large, it is consistent 
with the system ($Au+Au$ versus $d+Au$) and $\sqrt{s_{NN}}$ dependence 
coming from $p_T$ diffusion~\cite{Gyulassy:2000gk}. We note that even 
at the maximum RHIC energy the $y=0$ central $d+Au$ collisions exhibit 
$\sim 35\%$ maximum pion enhancement~\cite{Frawley:2004gj}. 
The final state partonic energy  loss was computed for 
a range of initial effective gluon rapidity 
densities $dN^g/dy=650-800$ as discussed in Section~II. 
Figure~2(a) clearly shows that the perturbative calculation 
presented here predicts a dominance of the inelastic 
jet interactions and net $\pi^0$ quenching at the 
intermediate RHIC energy.

Additional constraints for the energy loss calculations 
would arise from the measurement of a suppressed
double inclusive hadron production 
 \begin{eqnarray}         
R^{h_1h_2}_{AA}  = 
\frac{d N^{AA}({\bf b})}{dy_1dy_2d^2{\bf p}_{T1}d^2{\bf p}_{T2} } \big / 
\frac{ T_{AA}({\bf b}) \;d\sigma^{pp} } 
 {dy_1dy_2d^2{\bf p}_{T1}d^2{\bf p}_{T2}   }_{}  \;. && \nonumber \\ 
&&
\label{geomfact2} 
\end{eqnarray}
The cross sections in Eq.~(\ref{geomfact2}) can be evaluated as 
in~\cite{Qiu:2004da} by fragmenting the second hard scattered parton.
We note that by unitarity the details of its energy loss and 
fragmentation do not affect the calculation of the single inclusive
spectra. Di-hadron correlations, however, pick only one of the particle
states. Since the second parent parton also looses a fraction of 
its energy in the medium,  
qualitatively $ 1 \leq R^{h_1}_{AA} / R^{h_1h_2}_{AA} \leq 2$.  
Numerical estimates in Figure~2(b) from jet quenching alone 
indicate that the double inclusive 
hadron suppression is $25\% - 40\%$ larger than the single inclusive 
quenching in the $5 \leq p_{T1}= p_{T1} \leq 7 $~GeV range  at
$\sqrt{s_{NN}}=62$~GeV. The ratio was computed in the mean $\Delta E$ 
approximation, but with an explicit average over the di-jet production 
point for a realistic Woods-Saxon geometry. Double inclusive cross 
section quenching is manifest  in the attenuation of the away-side 
correlation function  $C_2(\Delta \phi) = (1/N_{\rm trig}) 
dN^{h_1h_2}/d\Delta \phi $~\cite{Qiu:2004da}.  
In Figure~2(c) it leads to a $3-5$ fold attenuation of the area 
$A_{\rm far}$ relative to the $p+p$ case. In contrast, elastic 
transverse momentum diffusion will only result in a 
broader $C_2(\Delta \phi)$.  We note that the experimentally 
measured suppression  value will be sensitive to the subtraction 
of the elliptic flow  $v_2$ component~\cite{Teaney:2000cw}, 
which may remove part or all of the 
residual correlations. The results shown in Figure~2(c), 
therefore, correspond to the smallest anticipated observable 
disappearance of the away-side jet at high $p_T$.

\section{Discussion and summary}

For $p_T \geq 2$~GeV the reduction of the pion 
multiplicity in central $Au+Au$ reactions relative 
to the binary collision scaled $p+p$ result at 
the intermediate RHIC energy of  $\sqrt{s_{NN}}=62$~GeV 
was found to be a factor of $2-3$. This result is  
consistent with $R_{AA}(p_T = 4\;{\rm GeV}) \sim 0.5$ 
from~\cite{Wang:2003aw}.
The moderate transverse momentum dependence of the quenching 
ratio arises from the interplay of the Cronin effect, 
in particular its decrease toward larger transverse momenta,  
and the jet energy loss. The medium induced energy loss  
also results in an apparent disappearance of the away-side
jet at high $p_T$, which is similar  to 
the $\sqrt{s_{NN}}=200$~GeV  result~\cite{Adler:2002tq}.

To relate the effective $dN^g/dy=650-800$ to the 
temperature  and energy density we employ an initial 
equilibration time $\tau_0 \simeq 0.8 - 1$~fm suggested 
by the hydrodynamic description of the SPS data~\cite{Teaney:2000cw}.
For 1+1D expansion  $\epsilon_0 = 6 - 8 $~GeV/fm$^3$, 
$T_0 = 300 - 325$~MeV and  the lifetime 
of the plasma phase $\tau-\tau_0 = 3.5 - 4$~fm. 
Such initial conditions are already significantly above 
the current expectation for the critical temperature
and  energy density  for a deconfinement 
phase transition, $\epsilon_c \simeq 1$~GeV and 
$T_c \simeq 175$~MeV~\cite{Karsch:2001vs}.

From a theoretical point of view, the most interesting 
result would be a strong deviation of the pion 
attenuation from the pQCD calculation in Figure~2(a). 
This  may be either a factor of $\sim 5$ suppression,  
comparable to the $\sqrt{s_{NN}}=200$~GeV quenching, 
or binary and above binary scaling, as seen at the SPS 
energies. Both cases would indicate a marked non-linear 
dependence of the non-Abelian energy loss on the soft parton 
rapidity density $dN^g/dy$ if we assume that parton-hadron
duality still holds. Equally striking will be an absence
of strong attenuation in the away-side 
$C_2(\Delta \phi)$.  It is important to check
whether the established connection between the single and 
double inclusive hadron suppression~\cite{Adler:2002tq}   
persist at the intermediate RHIC energies, see Figure 2(c).

In summary, the upcoming $\sqrt{s_{NN}}=62$~GeV 
measurements and their comparison to theoretical calculations
will provide a critical test of jet tomography. They will help 
to further clarify the properties of the hot and dense nuclear 
matter crated in relativistic nuclear collisions and the 
position that it occupies on the QCD phase transition diagram.

{\em Note added:} Shortly after the completion of this work 
preliminary  PHENIX data on $\pi_0$ production and attenuation 
at the intermediate RHIC energy of $\sqrt{s_{NN}}=62$~GeV became 
available. Comparison between the data and the predictions 
given in this letter can be found in~\cite{Awes}.

\begin{acknowledgments}
Useful discussion with J.~W.~Qiu  and M.~B.~Johnson is gratefully 
acknowledged. This work is supported  by the Director, Office of Science, 
Office of High Energy and Nuclear Physics, Division of Nuclear Physics 
of the US Department of Energy under Grant No. DE-FG02-87ER40371.
\end{acknowledgments}


\begin{thebibliography}{80}
\bibitem{quarks}
Hard jets have a $C_F/C_A = 4/9$  smaller elastic scattering 
cross section with quarks relative to gluons. If nuclear matter 
is approximated by effective gluons, the inferred density will 
be correspondingly smaller  while still yielding the same 
mean $\sigma_{el}$ and opacity.
\end{thebibliography}
\end{document}